# Multi-Bernoulli Mixture Filter: Complete Derivation and Sequential Monte Carlo Implementation

Sen Wang

*Abstract*—Multi-Bernoulli mixture (MBM) filter is one of the exact closed-form multi-target Bayes filters in the random finite sets (RFS) framework, which utilizes multi-Bernoulli mixture density as the multi-target conjugate prior. This filter is the variant of Poisson multi-Bernoulli mixture filter when the birth process is changed to a multi-Bernoulli RFS or a multi-Bernoulli mixture RFS from a Poisson RFS. On the other hand, labeled multi-Bernoulli mixture filter evolves to MBM filter when the label is discarded. In this letter, we provide a complete derivation of MBM filter where the derivation of update step does not use the probability generating functional. We also describe the sequential Monte Carlo implementation and adopt Gibbs sampling for truncating the MBM filtering density. Numerical simulation with a nonlinear measurement model shows that MBM filter outperforms the classical probability hypothesis density filter.

## I. INTRODUCTION

THE objective of multi-target tracking (MTT) is to jointly estimate the number of targets and their individual states from a sequence of measurements provided by sensing devices such as radar [1], sonar [2], or cameras [3]. Relevant MTT algorithms are joint probabilistic data association filter (JPDAF) [4], multiple hypothesis tracking (MHT) [5] and filters based on random finite sets (RFS) [6, 7].

MTT methods based on RFS strictly describe target birth, death, spawning, miss detection and clutters in MTT process, directly estimate number and state of targets, and even provide target tracks or trajectories. In the recent years, some approximations of multi-target Bayes filters have been proposed, such as the probability hypothesis density (PHD) filter [8-10], the cardinalized PHD (CPHD) filter [11, 12] and the multi-Bernoulli filter [13]. Since the multi-target conjugate prior densities have been introduced, several exact closed-form multi-target Bayes filters have been proposed, including generalized labeled multi-Bernoulli (GLMB) filter [14], labeled multi-Bernoulli mixture (LMBM) filter [15], and Poisson multi-Bernoulli mixture (PMBM) filter [16-18].

If the birth process is changed to a multi-Bernoulli RFS or a multi-Bernoulli mixture (MBM) RFS from a Poisson RFS, the PMBM filter will become the MBM filter. On the other hand, the LMBM filter will evolve to the MBM filter when the label is discarded. Up to now, Gaussian Mixture (GM) implementation of MBM filter was presented, which is only suitable for linear/Gaussian models [19]. In this letter, the multi-Bernoulli mixture RFS is described and the complete derivation of MBM filter is provided. It is noteworthy that the derivation of update step does not rely on the probability generating functional (PGFL). Then sequential Monte Carlo (SMC) implementation of MBM filter is presented. In addition, Gibbs sampling is adopted to find the finite number of global hypotheses. Numerical simulation with a nonlinear measurement model demonstrates the performance of the MBM filter.

## II. BACKGROUND

### A. Bayesian Filtering Recursion

Bayesian filtering recursion in the RFS framework consists of the update and prediction steps. Suppose $X$ and $Z$ represent the multitarget state and measurements at the current step respectively, $f(X)$ is the multitarget prior density, and $l(Z|X)$ is the multitarget likelihood function, then the multitarget posterior density is given as

$$q(X) = \frac{f(X)l(Z|X)}{\int f(X)l(Z|X)\delta X} \quad (1)$$

Suppose $X'$ represents the multitarget state at the next time step, and $\gamma(X'|X)$ is the multitarget Markov density, then the multitarget prior density at the next time step is given as

$$f(X') = \int \gamma(X'|X)q(X)\delta X \quad (2)$$

### B. Multi-Bernoulli Mixture RFS

An RFS is simply a finite-set-valued random variable. The disjoint union of a fixed number of independent Bernoulli RFSs produces a multi-Bernoulli RFS, and a weighted sum of multi-

Corresponding author: Sen Wang.
The author is with National Key Laboratory of Science and Technology on ATR, National University of Defense Technology, Changsha, China (e-mail: wangsen11@nudt.edu.cn).

Bernoulli RFSs produces a multi-Bernoulli mixture RFS. The MBM density is

$$f(X) = \sum_h w_h \sum_{X_1 \uplus \cdots \uplus X_n = X} \prod_{i=1}^n f_{h,i}(X_i) \quad (3)$$

where $w_h$ represents the weight of hypothesis $h$ and $\sum_h w_h = 1$, $\uplus$ denotes disjoint union, and $X_i$ represents the $i$th Bernoulli RFS and has density as

$$f_{h,i}(X_i) = \begin{cases} 1 - r_{h,i} & X_i = \varnothing \\ r_{h,i} p_{h,i}(x_i) & X_i = \{x_i\} \end{cases} \quad (4)$$

in hypothesis $h$, where $r_{h,i}$ represents the existence probability and $p_{h,i}$ represents the state density given that $X_i$ exists.

Suppose $g$ represents the test function, then the probability generating functional of (3) is

$$G[g] = \sum_h w_h \prod_{i=1}^n 1 - r_{h,i} + r_{h,i} \langle p_{h,i}, g \rangle \quad (5)$$

where $\langle a, b \rangle = \int a(x) b(x) dx$ represents the inner product of $a(x)$ and $b(x)$.

III. DERIVATION OF MBM FILTER

A. Derivation of Update Step

Suppose the multitarget prior density $f(X)$ is an MBM density of the form (3) and (4), the multitarget state and measurements are $X = X_1 \uplus \cdots \uplus X_n$ and $Z = \{z_1, \cdots, z_m\}$ respectively, in this section we prove that the multitarget posterior density is also an MBM density.

Due to the data association uncertainty, we must consider all possible hypotheses. An association function is defined as: $\theta: \{1, \cdots, n\} \to \{0, 1, \cdots, m\}$ such that $\theta(i) = \theta(i') > 0$ implies $i = i'$. $i \to \theta(i) \neq 0$ indicates target $X_i$ is associated with measurement $z_{\theta(i)}$, and the corresponding likelihood function is $l(z_{\theta(i)} | X_i)$. $i \to \theta(i) = 0$ indicates target $X_i$ is undetected, and the corresponding likelihood function is $l(\varnothing | X_i)$. The set of clutter measurements is defined as

$$Z^c = Z \setminus \{z_{\theta(i)} | \theta(i) > 0, i \in \{1, \cdots, n\}\} \quad (6)$$

where $A \setminus B$ returns elements in $A$ that are not in $B$. The set $Z^c$ is modeled as a Poisson RFS with the intensity $c(\cdot)$, and has the density function as

$$f(Z^c) = e^{-\int c(z) dz} [c(\cdot)]^{Z^c} \quad (7)$$

where $[f(\cdot)]^X = \prod_{x \in X} f(x)$ and $[f(\cdot)]^{\varnothing} = 1$ by definition. One $\theta$ corresponds to one association, and the multitarget likelihood function is

$$l(Z | X) = \sum_\theta \left\{ \prod_{i:\theta(i)>0} l(z_{\theta(i)} | X_i) \prod_{i:\theta(i)=0} l(\varnothing | X_i) f(Z^c) \right\} \quad (8)$$

Substituting (3) and (8) into (1), we can obtain the multitarget posterior density

$$q(X) = \frac{e^{-\int c(z) dz} \prod_{z \in Z} c(z)}{\int f(X) l(Z | X) \delta X} \times$$
$$\sum_h \sum_\theta \left\{ w_h \sum_{X_1 \uplus \cdots \uplus X_n = X} \left\{ \prod_{i:\theta(i)>0} \frac{f_{h,i}(X_i) l(z_{\theta(i)} | X_i)}{c(z_{\theta(i)})} \times \right. \right. \quad (9)$$
$$\left. \left. \prod_{i:\theta(i)=0} f_{h,i}(X_i) l(\varnothing | X_i) \right\} \right\}$$

For any detected target $i$,

$$\frac{f_{h,i}(X_i) l(z_{\theta(i)} | X_i)}{c(z_{\theta(i)})}$$
$$= \frac{r_{h,i} \int p_{h,i}(x_i) p_d(x_i) l(z_{\theta(i)} | x_i) dx_i}{c(z_{\theta(i)})} \times \quad (10)$$
$$\frac{p_{h,i}(x_i) p_d(x_i) l(z_{\theta(i)} | x_i)}{\int p_{h,i}(x_i) p_d(x_i) l(z_{\theta(i)} | x_i) dx_i}$$

indicates that target $X_i$ is a Bernoulli RFS whose existence probability is 1.

For any undetected target $i$,

$$f_{h,i}(X_i) l(\varnothing | X_i)$$
$$= \left\{ 1 - r_{h,i} + r_{h,i} \int p_{h,i}(x_i)(1 - p_d(x_i)) dx_i \right\} \times \quad (11)$$
$$\begin{cases} \dfrac{1 - r_{h,i}}{1 - r_{h,i} + r_{h,i} \int p_{h,i}(x_i)(1 - p_d(x_i)) dx_i} & X_i = \varnothing \\ \dfrac{r_{h,i} p_{h,i}(x_i)(1 - p_d(x_i))}{1 - r_{h,i} + r_{h,i} \int p_{h,i}(x_i)(1 - p_d(x_i)) dx_i} & X_i = \{x_i\} \end{cases}$$

indicates that target $X_i$ is also a Bernoulli RFS.

Consequently, the multitarget posterior density is also an MBM density, and association $\theta$ in prior hypothesis $h$ produces a posterior hypothesis with unnormalized weight

$$w_{h,\theta} = w_h \times \prod_{i:\theta(i)>0} \frac{r_{h,i} \int p_{h,i}(x_i) p_d(x_i) l(z_{\theta(i)} | x_i) dx_i}{c(z_{\theta(i)})} \times \quad (12)$$
$$\prod_{i:\theta(i)=0} \left\{ 1 - r_{h,i} + r_{h,i} \int p_{h,i}(x_i)(1 - p_d(x_i)) dx_i \right\}$$

B. Derivation of Prediction Step

Suppose the multitarget posterior density $q(X)$ is an MBM density of the form (3) and (4), and the birth density is modeled as a multi-Bernoulli density, in this section we use PGFL to prove that the multitarget prior density at the next time step is also an MBM density.

If the set of birth targets is a multi-Bernoulli RFS with $n^b$ components, and each one has existence probability $r_i^b$ and state density $p_i^b$, the PGFL of birth density is

$$G_B[g] = \prod_{i=1}^{n^b} 1 - r_i^b + r_i^b \langle p_i^b, g \rangle \quad (13)$$

Suppose $p_s(x)$ is the probability that single target $x$ at the current step will survive at the next time step, and $f(x' | x)$ is

the single-target Markov transition density, then the PGFL of (2) is [6, p. 682]

$$G_{X'}[g] = G_B[g]G[1 - p_s + p_s p_g]$$ (14)

where $p_g(x) = \langle f(\cdot|x), g \rangle$.

Substituting (5) and (13) into (14), we can obtain

$$G_{X'}[g] = \sum_h \left\{ w_h \prod_{i=1}^{n^b} 1 - r_i^b + r_i^b \langle p_i^b, g \rangle \times \right.$$

$$\left. \prod_{i=1}^{n} 1 - r_{h,i} \langle p_{h,i}, p_s \rangle + r_{h,i} \langle p_{h,i}, p_s \rangle \left\langle \frac{\int p_{h,i}(x) p_s(x) f(\cdot|x) dx}{\langle p_{h,i}, p_s \rangle}, g \right\rangle \right\}$$ (15)

which indicates the multitarget prior density at the next time step is also an MBM density with the same hypothesis weight as that of the multitarget posterior density at the current step.

If the birth density is modeled as an MBM density, the multitarget prior density at the next time step is still an MBM density.

## IV. IMPLEMENTATION OF MBM FILTER

In this section, SMC implementation is provided, and Gibbs sampling truncates MBM density. Furthermore, target state estimation and target/hypothesis pruning are also discussed.

### A. Implementation of Update Step

In the update step, each prior hypothesis grows up to multiple posterior hypotheses without changing the number of targets.

Prior target $X_i$ in prior hypothesis $h$ is described by the existence probability $r_{h,i}$ and $n^p$ particles with equal weights: $\{x_i^{(p)}, 1/n^p\}_{p=1}^{n^p}$, and the latter is the approximation of state density $p_{h,i}$.

In association $\theta$, if measurement $z_{\theta(i)}$ is associated with $X_i$, the posterior existence probability is 1, and the posterior state density can be described by the following particles:

$$\left\{ x_i^{(p)}, \frac{p_d(x_i^{(p)}) l(z_{\theta(i)} | x_i^{(p)})}{\sum_{p=1}^{n^p} p_d(x_i^{(p)}) l(z_{\theta(i)} | x_i^{(p)})} \right\}_{p=1}^{n^p}$$ (16)

The contribution of target $X_i$ to posterior hypothesis weight $w_{h,\theta}$ is

$$C_i = \frac{r_{h,i} \sum_{p=1}^{n^p} p_d(x_i^{(p)}) l(z_{\theta(i)} | x_i^{(p)})}{n^p c(z_{\theta(i)})}$$ (17)

In association $\theta$, if target $X_i$ is undetected, the posterior existence probability is

$$\frac{r_{h,i}/n^p \sum_{p=1}^{n^p} 1 - p_d(x_i^{(p)})}{1 - r_{h,i} + r_{h,i}/n^p \sum_{p=1}^{n^p} 1 - p_d(x_i^{(p)})}$$ (18)

and the posterior state density can be described by the following particles:

$$\left\{ x_i^{(p)}, \frac{1 - p_d(x_i^{(p)})}{\sum_{p=1}^{n^p} 1 - p_d(x_i^{(p)})} \right\}_{p=1}^{n^p}$$ (19)

The contribution of target $X_i$ to posterior hypothesis weight $w_{h,\theta}$ is

$$C_i = 1 - r_{h,i} + r_{h,i}/n^p \sum_{p=1}^{n^p} 1 - p_d(x_i^{(p)})$$ (20)

Particle weights in (16) and (19) are not equal, and the resampling technique can be utilized to replace them with new, equal weights.

According to (12), unnormalized posterior hypothesis weight $w_{h,\theta}$ is the product of prior hypothesis weight $w_h$ and the contribution of each target $C_i$. Normalized posterior hypothesis weight is $w_{h,\theta} / \sum_h \sum_\theta w_{h,\theta}$.

### B. Implementation of Prediction Step

In the prediction step, the number of hypotheses doesn't change while the number of targets in each hypothesis increases by $n^b$. That is to say, each hypothesis is augmented with the Bernoulli components of birth targets.

Similarly to IV-A, posterior target $X_i$ in posterior hypothesis $h$ is described by $r_{h,i}$ and $\{x_i^{(p)}, 1/n^p\}_{p=1}^{n^p}$.

At the next time step, the prior existence probability of surviving target $X_i$ is

$$r_{h,i}/n^p \sum_{p=1}^{n^p} p_s(x_i^{(p)})$$ (21)

and the prior state density can be described by the following particles:

$$\left\{ x_i'^{(p)}, \frac{p_s(x_i^{(p)})}{\sum_{p=1}^{n^p} p_s(x_i^{(p)})} \right\}_{p=1}^{n^p}$$ (22)

where $x_i'^{(p)}$ is sampled from single-target Markov transition density $f(x'|x_i^{(p)})$. Particle weights in (22) are not equal, and the resampling technique can be utilized to replace them with new, equal weights.

The prior existence probability of $i$th birth target is $r_i^b$, and its state density can be described by the following particles:

$$\{x_i^{b(p)}, 1/n^p\}_{p=1}^{n^p}$$ (23)

where $x_i^{b(p)}$ is sampled from $p_i^b$.

## C. Gibbs Sampling

This letter approximates the update step by pruning the posterior hypotheses using Gibbs sampling. For each prior hypothesis $h$, we use a $n$-tuple $\Theta = [\theta(1), \cdots, \theta(n)]$ to represent the association between targets and measurements, and we sample $k = \lceil N_h w_h \rceil$ posterior hypotheses from the distribution of $\Theta$, where $N_h$ is the maximum number of posterior hypotheses. If we have known the association $\Theta_{-i} = [\theta(1), \cdots, \theta(i-1), \theta(i+1), \cdots, \theta(n)]$, the possible value of $\theta(i)$ is the element in $\{0, 1, \cdots, m\} \setminus \Theta_{-i} \cup \{0\}$, and the conditional probability is proportional to the contribution of target $X_i$ to posterior hypothesis weight $w_{h,\theta}$

$$f(\theta(i) | \Theta_{-i}) \propto C_i \qquad (24)$$

## D. Estimation and Pruning

Suppose the multitarget posterior density is the form of (3) and (4), the hypothesis with maximum weight is selected

$$h^* = \arg\max_h w_h \qquad (25)$$

and target with existence probability $r_{h^*,i} > r^{th}$ is extracted, where $r^{th}$ is the threshold.

In order to maintain the same number of targets in all hypotheses, target $X_i$ with $\sum_h w_h r_{h,i} < r^p$ is pruned in all posterior hypotheses, where $r^p$ is the pruning threshold of targets. Posterior hypothesis $h$ with $w_h < w^p$ is pruned, and $w^p$ is the pruning threshold of hypotheses.

## V. SIMULATION

In this section, we use computer simulations to demonstrate the performance of the MBM filter. Sensor field of view is a 2-dimensional region $[-50, 50] \times [0, 100]$. The state equation and the measurement equation of single target are

$$x' = \left(\begin{bmatrix} 1 & 0 \\ 0 & 1 \end{bmatrix} \otimes \begin{bmatrix} 1 & T \\ 0 & 1 \end{bmatrix}\right) x + \begin{bmatrix} T^2/2 & T & 0 & 0 \\ 0 & 0 & T^2/2 & T \end{bmatrix}^T n \qquad (26)$$

$$z = \begin{bmatrix} \sqrt{p_x^2 + p_y^2} \\ \arctan(p_y/p_x) \end{bmatrix} + w \qquad (27)$$

where 4-dimensional target state $x$ and $x'$ consist of the target position and velocity along the x-axis and y-axis, i.e., $x = [p_x, v_x, p_y, v_y]^T$, $\otimes$ represents the Kronecker product, the bearing and the range are measured represented as $z$, sampling period $T = 1$, process noise $n \sim N(0, diag[4 \times 10^{-6}, 4 \times 10^{-6}])$, and measurement noise $w \sim N(0, diag[0.25, 0.09])$. This letter considers five targets with motion parameters showed in TABLE I, and the total time of simulation is 100.

The performance of MBM filter is evaluated in comparison with PHD filter. Parameters in MBM filter are as follows: the maximum number of posterior hypotheses $N_h = 100$, the pruning threshold of targets $r^p = 10^{-5}$, the pruning threshold of hypotheses $w^p = 10^{-5}$, target extraction threshold $r^{th} = 0.5$, the existence probability of birth targets $r_i^b = 0.01$ and the corresponding particles are sampled from single-target Markov transition density conditional on target initial state. In PHD filter, the probability density of birth targets is modeled as Gaussian mixture, which has the same particles as MBM filter and all weights are $10^{-5}$. In both filters, survival probability is $p_s = 0.99$ and $n^p = 10^3$ particles are used for per target.

TABLE I
MOTION PARAMETERS OF TARGETS

| Target | Initial state | Birth time | Death time |
|---|---|---|---|
| 1 | $[-50, 1.65, 100, -1.65]^T$ | 1 | 60 |
| 2 | $[-50, 1.65, 0, 1.65]^T$ | 11 | 70 |
| 3 | $[-50, 0.875, 30, 0.875]^T$ | 11 | 90 |
| 4 | $[50, -1.16, 70, -1.16]^T$ | 31 | 90 |
| 5 | $[50, -1.65, 50, 0]^T$ | 41 | 100 |

Poisson clutter is uniform in sensor field of view with an average intensity of $5 \times 10^{-4}$ and detection probability is set as $p_d = 0.9$. Average Optimal Sub-Pattern Assignment (OSPA) [20] of 100 Monte-Carlo simulations is used to evaluate the tracking performance, as depicted in Fig. 1, where the cut-off factor is 10 and the order is 2. Both filters can track targets effectively although larger errors occur at target birth time and death time. Furthermore, MBM filter is superior than PHD filter at most scans.

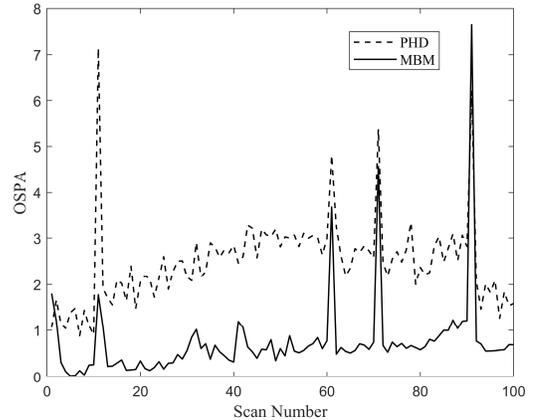

Fig. 1. Average OSPA of 100 Monte-Carlo simulations

## VI. CONCLUSION

In this letter, the complete derivation of MBM filter is proposed, where the derivation of prediction step relies on the PGFL while that of update step utilizes multi-target calculus. SMC implementation of MBM filter is presented and several strategies including Gibbs sampling and target/hypothesis pruning are used to improve computing efficiency. Finally, numerical simulation with a nonlinear measurement model demonstrates that the proposed filter outperforms PHD filter.